\documentclass[12pt]{article}

\usepackage{color}
\usepackage{amsmath}
\usepackage{amsfonts}
\usepackage{amssymb}
\usepackage{caption}
\usepackage{graphicx}
\usepackage{slashed}            
\usepackage{subfig}             
\usepackage{xspace}				
\usepackage{cite}

\usepackage[english]{babel}
\usepackage{fancyhdr}
\usepackage{amsmath}
\usepackage{amssymb}
\usepackage{amsfonts}
\usepackage{psfrag}
\usepackage[applemac]{inputenc}
\usepackage{graphicx}

\newcommand{\arXiv}[2]{\href{http://arxiv.org/pdf/#1}{{\tt #2/#1}}}
\newcommand{\arXivold}[1]{\href{http://arxiv.org/pdf/#1}{{\tt #1}}}

\renewcommand{\tilde}{\widetilde} 

\newcommand{\beq}{\begin{eqnarray}}
\newcommand{\eeq}{\end{eqnarray}}

\newcommand{\eq}[1]{Eq.~(\ref{#1})}

\newcommand{\bag}{\begin{align}}
\newcommand{\eag}{\end{align}}

\newcommand{\order}[1]{O(#1)}

\newcommand{\vev}[1]{\langle {#1} \rangle}

\newcommand{\nn}{\nonumber}

\newcommand{\Lag}{\mathcal{L}}

\usepackage[hypertexnames=false]{hyperref}		

\begin{document}
\begin{titlepage}

\vskip.5cm

\begin{center} 
{\huge \bf A Naturally Light Dilaton and a Small Cosmological Constant  \\ \vspace*{0.3cm} } 
\end{center}

\begin{center} 
{\bf \  Brando Bellazzini$^{a, b}$, Csaba Cs\'aki$^c$, Jay Hubisz$^d$,  Javi Serra$^c$, John Terning$^e$} 
\end{center}

\begin{center} 
$^{a}$ {\it  Dipartimento di Fisica e Astronomia, Universit\`a di Padova and INFN, Sezione di Padova, Via Marzolo 8, I-35131 Padova, Italy} \\

\vspace*{0.1cm}

$^{b}$ {\it SISSA, Via Bonomea 265, I-34136 Trieste, Italy} \\

\vspace*{0.1cm}

$^{c}$ {\it Department of Physics, LEPP, Cornell University, Ithaca, NY 14853, USA} \\

\vspace*{0.1cm}

$^{d}$ {\it  Department of Physics, Syracuse University, Syracuse, NY  13244} \\

\vspace*{0.1cm}

$^{e}$ {\it Department of Physics, University of California, Davis, CA 95616} \\

\vspace*{0.1cm}

{\tt  
 \href{mailto:brando.bellazzini@pd.infn.it}{brando.bellazzini@pd.infn.it},
\href{mailto:csaki@cornell.edu}{csaki@cornell.edu}, \\
 \href{mailto:jhubisz@phy.syr.edu}{jhubisz@phy.syr.edu},  
 \href{mailto:js993@cornell.edu}{js993@cornell.edu},
 \href{mailto:jterning@gmail.com}{jterning@gmail.com}}

\end{center}

\vglue 0.3truecm

\centerline{\large\bf Abstract}
\begin{quote}
We present a non-supersymmetric theory with a naturally light dilaton. It is based on a 5D holographic description of a conformal theory perturbed by a close-to-marginal operator of dimension $4-\epsilon$, which develops a condensate. As long as the dimension of the perturbing operator remains very close to marginal (even for large couplings) a stable minimum at hierarchically small scales is achieved, where the dilaton mass squared is suppressed by $\epsilon$. At the same time the cosmological constant in this sector is also suppressed by $\epsilon$, and thus parametrically smaller than in a broken SUSY theory. As a byproduct we also present an exact solution to the scalar-gravity system that can be interpreted as a new holographic realization of spontaneously broken conformal symmetry. Even though this  metric deviates substantially from AdS space in the deep IR it still describes a non-linearly realized exactly conformal theory. We also display the effective potential for the dilaton for arbitrary holographic backgrounds.

\end{quote}

\end{titlepage}


\setcounter{equation}{0}
\setcounter{footnote}{0}

\section{Introduction}

Dynamical spontaneous breaking of scale invariance (SBSI) is rare. If a theory is exactly conformal, it either does not break scale invariance, or the breaking scale  is arbitrary (a flat direction)~\cite{Fubini:1976jm}.  Thus an explicit breaking has to be present to trigger and stabilize the SBSI. However, this explicit breaking must remain small throughout the whole renormalization group running not to ruin scale invariance. In particular, the $\beta$-function of the coupling that introduces the explicit breaking must remain small at the scale of SBSI. This condition is difficult to satisfy: for example in QCD or technicolor (TC) models the condensates are triggered by large and rapidly changing couplings at the condensation scale $\Lambda_{QCD,TC}$, implying large explicit breaking. Thus no light dilaton is expected in either case (in agreement with the absence of a light dilaton-like scalar hadron in QCD)~\cite{Holdom}. 

One possible scenario is that conformality is spontaneously broken along a flat direction, which is then lifted via the potential generated through a small external coupling whose non-zero but also small $\beta$-function breaks scale invariance explicitly.  This mechanism is essentially what is assumed to happen in the Randall-Sundrum (RS) model~\cite{RS} stabilized via the  Goldberger-Wise (GW) mechanism~\cite{GW}: the bulk scalar field is associated with the small and slowly running external coupling, and the appearance of the IR brane signals that SBSI has occurred~\cite{RZ,APR}. The resulting massive radion mode~\cite{CGRT,GW2,Csaki:2000zn,Csaki:2007ns} is identified with the light dilaton of SBSI.  However, in the 5D picture this scenario assumes that the IR brane tension has been tuned such that the radion potential is flat in the absence of the external perturbation. Generically such fine tunings are not  present, and one would like to understand whether a light dilaton can still occur in the absence of tuning. In the 4D language the theory with a large mistune can be understood in terms of the dilaton potential in the following way~\cite{ourdilaton,chackodilaton}.  Scale invariance allows an unsuppressed quartic (non-derivative) self-interaction term for the dilaton, since a dimension four operator in the action is scale invariant. An $\order{1}$ mistune on the IR brane corresponds to a large quartic dilaton potential, which would generically prevent SBSI, at least for small perturbations. 
In the 5D picture a non-vanishing quartic would force the IR brane to infinity (and thus no SBSI), or the branes would be very close to each other (so effectively no scale invariant regime).

Contino, Pomarol and Rattazzi (CPR) have suggested in an important unpublished work~\cite{CPR} that this might be overcome if the quartic becomes mildly energy dependent via an explicit scale-invariance breaking perturbation, whose $\beta$-function remains parametrically small, but not necessarily the coupling itself. In this case the expectation is that SBSI will happen around the scale where the effective dilaton quartic vanishes, which can be a hierarchically small scale if the running lasts for a long time. At the same time the dilaton can be light, if the $\beta$-function is parametrically small at the scale of spontaneous breaking. 
The latter is the crucial dynamical assumption: the perturbation, which might start small in the UV, becomes sufficiently large in the IR to neutralize the initial large quartic, but at the same time its $\beta$-function must remain small.
Phrased in a different way, as long as the $\beta$-function of the perturbation remains small, SBSI will naturally happen regardless of the absence of a flat direction to start with.

The mechanism of scanning through the possible values of the dilaton quartic coupling until it reaches the minimum where it  almost vanishes  is also important when we couple the theory to gravity. The value of the potential at the minimum corresponds to the cosmological constant contribution generated during the phase transition from a scale invariant theory to the broken phase. If the value of the potential at the minimum is naturally suppressed by the smallness of the $\beta$-function in the IR, the contribution to the cosmological constant could be significantly reduced. This is a mechanism along the lines Weinberg was considering in~\cite{Weinberg}, except he was requiring that the cosmological constant vanishes exactly, which in turn requires an exactly vanishing $\beta$-function. However in this case no dilaton stabilization can happen. The potential significance of the dilaton for reducing the cosmological constant was also emphasized in~\cite{Sundrum}. 

The aim of this paper is to examine the CPR proposal in a holographic setting and establish that it can indeed be a viable route towards finding a parametrically light dilaton in a dynamical SBSI theory with hierarchical scales. 
We will argue that even though the metric can deviate significantly from AdS space, this is due to the formation of a condensate of the perturbing operator which is very close to dimension four. As long as the dimension is very close to marginal ($4-\epsilon$), the condensate will correspond to pure spontaneous breaking of scale invariance, with the resulting contribution to the dilaton potential still corresponding to a  quartic (which will however acquire a mild scale dependence due to the running, $\epsilon \neq 0$). Moreover, as long as the running is slow, $\epsilon \ll 1$, the condensation in the IR will be universal: it will not depend on the details of the exact form of the $\beta$-function (which is captured by the form of the bulk scalar potential). Therefore in the IR the solutions to the coupled tensor-scalar equations will be well-approximated by the exact solution to the system with a dimension four condensate (corresponding to the case of no bulk scalar potential aside from the negative cosmological constant). In the UV a slowly running solution perturbing the AdS background can be used. These solutions can be joined  using asymptotic matching.\footnote{This matching was also recently used  by Chacko, Mishra and Stolarski~\cite{Chacko} for finding the solution to a particular bulk potential providing small perturbations around AdS.} This way we will be able to explicitly calculate the effective dilaton potential and show that the mass is suppressed by the small parameter $\epsilon$ of the $\beta$-function at the minimum of the potential. This yields an explicit construction for a dilaton that is parametrically lighter than the dynamical scale of the theory as required for models where the dilaton is a Higgs-like particle~\cite{ourdilaton,chackodilaton,Kitano,Witek}. Moreover, we show that the value of the dilaton potential at the minimum, which provides the cosmological constant contribution from the phase transition, is also suppressed by $\epsilon$. On the way we present an exact solution to the scalar-gravity system which is the gravity dual of a dimension 4 operator condensing in the IR, thereby yielding a fully spontaneous breaking of scale invariance. Even though the scalar background is not flat, and the deviation of the metric from AdS is large in the IR, this theory still realizes an exactly conformal theory that is spontaneously broken.  

CPR also comment on the nature of the bulk scalar and the origin of its suppressed potential: a large coupling with a small  $\beta$-function in 4D may be dual of a 5D Goldstone boson of the bulk with the potential suppressed by the Goldstone shift symmetry.  Of course, other realizations of small $\beta$-functions can be envisioned as well, e.g. the coupling approaching a strongly interacting IR fixed point that is not reached because of \textit{early} condensation. 
The construction presented here can be thought of as the proper realization of walking in technicolor theories \cite{walking}: in order to obtain a light dilaton the $\beta$-function needs to remain small even at the scale where the condensates are generated.  In the following we will not actually need to commit to any specific realization and the only crucial assumption is that the bulk potential is suppressed by a small symmetry breaking parameter.

The paper is organized as follows: in Sec.~\ref{sec:overview} we give an overview of the mechanism for obtaining a light dilaton and in particular emphasize the differences between the standard GW picture and the CPR proposal. In Sec.~\ref{sec:effpotential} we show how to calculate the dilaton effective potential in general holographic theories where the metric could deviate from AdS significantly. Sec.~\ref{sec:nobulkmass} is devoted to the discussion of the solution with a dimension 4 condensate (vanishing bulk scalar mass), and how to obtain a flat dilaton potential in that case via tuning two condensates against each other. Finally in Sec.~\ref{sec:CPR} we show how a naturally light dilaton can be obtained via the introduction of the small bulk mass, and comment on the suppression of the resulting cosmological constant in that case. Several appendices are devoted to alternative derivation of the dilaton effective potential (\ref{App:GH}), the detailed derivation of the small back-reaction case (\ref{App:smallback}) and the GW case (\ref{App:GW}), an explanation of the asymptotic matching procedure for the boundary layer problem used for finding the full solution (\ref{App:Matching}), a discussion of the dilaton kinetic term as well as dilaton parametrizations (\ref{App:kinetic}), and finally a discussion on an alternative choice for the IR brane potential (\ref{App:Linearpotential}).


\section{Light dilatons via long running and small $\beta$-function\label{sec:overview}}
\setcounter{equation}{0}
\setcounter{footnote}{0}

Unlike for internal symmetries, non-linearly realized spontaneously broken scale invariance allows a non-derivative quartic self-interaction for the dilaton:
\begin{equation}
V_{eff} = F \chi^4
\end{equation}
where $\chi$ is the dilaton field with scaling dimension one. For a theory without explicit breaking one needs to have $F=0$ in order for SBSI to occur: if $F>0$, the minimum is at $\chi=0$ (no SBSI), while for $F<0$ we find $\chi\to\infty$, thus there is no scale invariant theory. So the only possibility is that $F=0$, and thus $\chi$ is a flat direction: just like the flat potential valley for ordinary Goldstone bosons, the main difference being that the dilaton corresponds to a non-compact flat direction. If one wants to stabilize the scale one needs to introduce a small explicit breaking by perturbing the theory with a close-to-marginal operator ${\cal O}$ with a slowly running coupling $\lambda$. This will generate a small non-trivial potential 
\begin{equation}
V_{eff} = \chi^4 F(\lambda (\chi )) \ , \quad F(\lambda=0) \sim 0
\label{eq:Veff}
\end{equation}
which can introduce a non-trivial minimum for the potential at hierarchically small dilaton values, and give rise to a small dilaton mass.

$F=0$ and the appearance of a flat direction is natural in supersymmetric theories. 
Focusing on non-supersymmetric theories, one may ask how likely it is for $F\sim 0$ to occur in any given theory. The simplest answer is to perform an NDA analysis in the low-energy effective theory for the dilaton  which gives an estimate for the size of the quartic~\cite{ourdilaton} $F\sim 16 \pi^2$. From this point of view spontaneous scale symmetry breaking looks quite unlikely and tuned at best in non-SUSY theories. This issue is even more evident if we notice that by reparametrizing the dilaton as $\chi=fe^{\sigma/f}$ with $\vev{\sigma}=0$, the question of $F=0$ is reminiscent of a vanishing cosmological constant, $\Lambda_{eff} = F f^4$. 

\begin{figure}[!t]
\begin{center}
\includegraphics[width=4.0in]{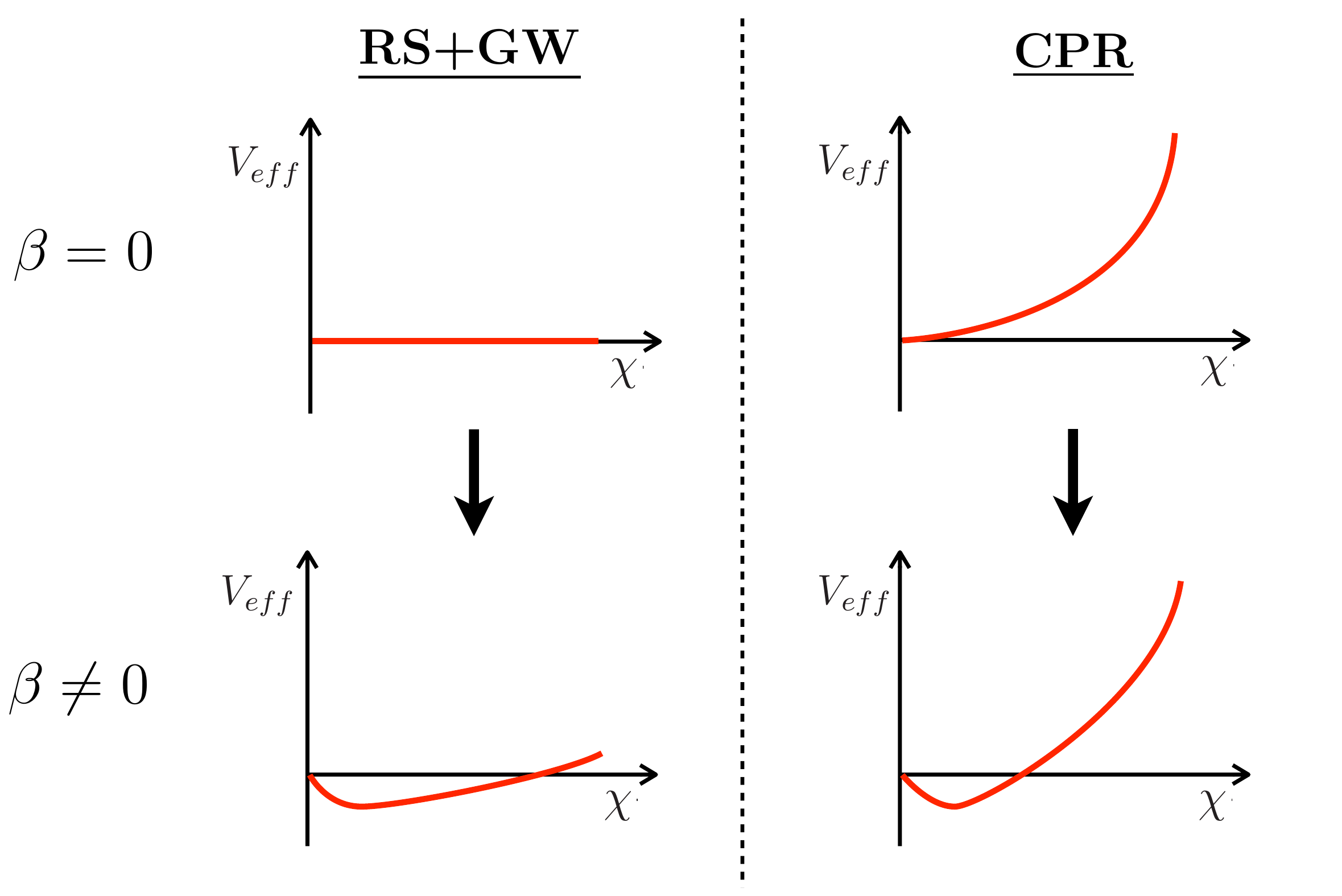}
\caption{Pictorial representation of the tuned scenario with vanishing quartic in the absence of stabilizing perturbation (left) versus the proposal discussed in this work, where a large perturbation compensates for the large initial quartic (right).}
\label{potentials}
\end{center}
\end{figure}

Contino, Pomarol and Rattazzi~\cite{CPR} have however suggested a different viewpoint: the presence of a flat direction  (in the absence of perturbation)  is not required (nor is it natural). Their approach is then that a theory with $F\neq 0$ will simply not break scale invariance spontaneously. Thus for a successful breaking of scale invariance a theory needs to be able to scan its value of $F$, until $F\sim 0$ is reached. In effect one needs a scale dependent quartic $F(\mu )$, which can be achieved by introducing  an external coupling $\lambda$, explicitly breaking scale invariance via its running
\begin{equation}
\frac{d\lambda}{d\log \mu } = \beta (\mu ) \equiv \epsilon \, b(\lambda) \ll 1\ ,
\end{equation}
where $b(\lambda)$ is a generic function of $\lambda$, whose detailed form is not important as long as the small parameter $\epsilon$ can be factored out.
This running coupling will in effect adjust the value of $F$ from its UV value (presumably of order $\sim 16 \pi^2$). If sufficiently long running is allowed, the corrections $\delta F \sim (\Lambda_{UV}/\mu)^\epsilon$ can become sizable, and at some scale $\mu_{IR}$ we find $F(\lambda(\mu_{IR}) ) \sim 0$. At this scale spontaneous breaking of scale invariance can happen. 
Since scale invariance is effectively recovered by substituting $\mu \to \chi$, this mechanism is equivalent to a generation of a non-trivial potential for the dilaton, \eq{eq:Veff}, but with $F(\lambda = 0 ) \sim 16 \pi^2$, and with its minimum determined by $F(\lambda(\chi)) \sim 0$.
Thus the CPR idea is to let the theory scan through the values of $F$ driven by the small explicit breaking term. The running will stop when the critical value $F \sim 0$ is reached and spontaneous breaking of scale invariance will occur. The differences between the scenario with $F \sim 0$, to which we refer as RS+GW (recalling its extra-dimensional realization), and $F \sim 16 \pi^2$, are illustrated in Fig.~\ref{potentials}. It is of course very important that the explicit breaking of scale invariance, that is the $\beta$-function, remains very small all throughout the running, and in particular at the IR scale where $F \sim 0$, otherwise the dilaton would pick up a large mass. This is exactly what happens in QCD or in technicolor: one starts out with a small $\beta$-function and an approximately conformal theory in the UV. However, in the IR the coupling and $\beta$ become large, and thus at energies where the QCD (or techniquark) condensates form there is no longer an approximate scale invariance and hence no light dilaton is expected, in accordance with the absence of an additional light scalar in QCD.

In order for the scanning mechanism to be possible, the contribution of the perturbation must approximately cancel the existing large tree-level quartic in the dilaton potential. This can happen only if the value of the coupling of the perturbing operator eventually becomes large. That does not automatically imply a large dilaton mass as long as the $\beta$-function remains small even while the coupling, $\lambda$, itself is big. This cancelation can be understood as follows: the increase in $\lambda$ along the running will be accompanied by a condensate for the perturbing operator ${\cal O}$, which will contribute a term $\propto \chi^{4-\epsilon}$ to the dilaton potential. If $\epsilon$ is very small (that is the condensate $\langle {\cal O}\rangle$  is very close to dimension four) this term can cancel the existing tree-level quartic at a hierarchically smaller scale than where the running started, and allow the CFT operators to also condense, giving rise to SBSI. Thus in this case the external perturbation both triggers and stabilizes SBSI. One can see that for this to work it is essential for ${\cal O}$ to be very close to dimension four, that is $\epsilon \ll 1$ throughout the running and even when $\lambda$ becomes sizable at the condensation scale.
 
The issue of whether a theory can scan through the possible values of $F$ and settle at a minimum where $F\sim 0$ is particularly interesting since in a gravitational theory the value of $F$ at the minimum  corresponds to the cosmological constant generated during the phase transition from the scale invariant to the broken phase. If $F\sim 0$ is natural in a theory coupled to gravity then a large (and perhaps most problematic) part of the total cosmological constant could be significantly reduced. The full cosmological constant in a model with an approximately conformal sector giving rise to electroweak symmetry breaking is made up of
\begin{equation}
V_{tot}=V_{UV}+V_{TeV}+V_{IR}
\end{equation}
where $V_{UV}$ is the value of the cosmological constant at the UV cutoff scale, $V_{TeV}$ is the contribution of the broken conformal sector (generically expected to be of the size $({\rm TeV})^4$ and contains the contribution from electroweak symmetry breaking), while $V_{IR}$ is the contribution due to all low-scale phase transitions well below the electroweak scale (for example due to the QCD phase transition). In a holographic model $V_{UV}$ would correspond to the contribution of the UV degrees of freedom localized on or around the UV brane, $V_{TeV}$ that of the degrees of freedom localized on or around the IR brane, while $V_{IR}$ is the contribution from physics below the scale of the lightest bulk KK mode or radion mode, where the theory effectively becomes four dimensional. One could perhaps imagine eliminating $V_{UV}$ via high scale SUSY, with a non-trivial interplay between SUSY and the conformal 
symmetry (for example SUSY might only be broken on the IR brane). Another possibility would be to use the hierarchically small dilaton VEV only to solve a little hierarchy between 10 TeV and 1 TeV, while above 10 TeV the theory becomes supersymmetric. 
In the model with dynamical SBSI triggered by the field of dimension $4-\epsilon$ discussed above, the expectation is that the value of the minimum of the potential will be suppressed by $\epsilon$ (since for $\epsilon \to 0$ the entire potential is again a pure quartic that forces $\chi = 0$), thus 
\begin{equation}
V_{TeV} \sim \epsilon ({\rm TeV})^4\ .
\label{eq:ccvalue}
\end{equation}
Finally, the contribution of the IR scale physics is expected to be of order  $V_{IR} \sim m_{dil}^4/(16 \pi^2)$, set by the size of the dilaton mass. If QCD was composite and the dilaton mass is smaller than the QCD scale then the energy from the QCD phase transition would be incorporated to a  contribution to $F$, and already be part of the estimate in (\ref{eq:ccvalue}). If the dilaton mass is heavier than the QCD scale then there will already be loop contributions to the cosmological constant in the 4D theory above the QCD scale which will be the leading contributions to $V_{IR}$. The dilaton mass (as we will see later) is expected to scale with $\epsilon$ as $m_{dil}^2 \propto \epsilon ({\rm TeV})^2$, thus the leading cosmological constant is given by (\ref{eq:ccvalue}). In order to reduce this to observed magnitudes one would need $\epsilon \lesssim 10^{-60}$. The associated approximately massless dilaton would mediate a long range force similar to gravity, with strength $\propto 1/({\rm TeV})^2$ \cite{Sundrum}. Fifth force bounds require that $\epsilon \gtrsim 6 \times 10^{-12}$ \cite{Lamoreaux:1996wh,Mostepanenko:2008vb} (corresponding to $m_{dil}\gtrsim 6$ eV).
The conservative option is then to assume that \eq{eq:ccvalue} is tuned against $V_{UV}$ to yield the observed cosmological constant.
 
Weinberg has argued~\cite{Weinberg} that a dilaton-like field can not be used to relax the cosmological constant to zero: if the theory is exactly conformal ($\epsilon =0$) the dilaton does not get stabilized, and one needs tuning to set the cosmological constant to zero, while for a broken theory ($\epsilon \neq 0$) the cosmological constant is not zero. This is not in contradiction with the arguments here. We will indeed see that for $\epsilon = 0$ one needs to tune the parameters of the theory in order to obtain a flat dilaton (and a vanishing low-energy contribution to the cosmological constant). However, we will see that for $\epsilon \neq 0$ the theory can relax to a vacuum with a small ($\epsilon$-suppressed) vacuum energy.

\section{The dilaton effective potential in holographic\\ models\label{sec:effpotential}}
\setcounter{equation}{0}

A general holographic model can be obtained by considering the action 
\begin{equation}
S = \int d^5 x \sqrt{g} \left( -  \frac{1}{2 \kappa^2} \mathcal{R} + \frac{1}{2} g^{MN} \partial_M \phi \partial_N \phi - V(\phi) \right)-\int d^4x \sqrt{g_0} V_0 (\phi) - \int d^4x \sqrt{g_1} V_1 (\phi).
\label{eq:action}
\end{equation}
of a bulk scalar field $\phi$ coupled to gravity. Here  $\kappa^2$ is the 5D Newton constant, which is related to 5D Planck scale via $\kappa^2 = \frac{1}{2 M_*^3}$. We will be considering 4D Lorentz invariant solutions to the Einstein equations, thus our metric ansatz will be 
\begin{equation}
ds^2 = e^{-2 A(y)} dx^2 - dy^2.
\label{eq:ansatz}
\end{equation}
where $e^{-A(y)}$ is the general warp factor. The AdS/CFT prescription gives an identification between the extra dimensional coordinate and an energy scale in a dual 4D CFT:  
\begin{equation}
\mu 
=k e^{-A(y)} \ ,
\end{equation}
where 
$k = \sqrt{ \frac{-\Lambda_{(5)}\kappa^2}{6} }$ is the curvature of the AdS space, determined by the 5D cosmological constant $\Lambda_{(5)}$.

We can then calculate the effective potential for the dilaton for an arbitrary background. We will assume that the general background is cut off at the position $y=y_1$ with orbifold boundary conditions, which corresponds to the presumed spontaneous breaking of conformality. 
The dilaton is identified as the scale of the spontaneous breaking, which in this case corresponds to the IR brane position $y_1$, implying
\begin{equation}
\chi =  e^\frac{\sigma}{f} = e^{-A(y_1)}\ .
\label{eq:holodila}
\end{equation}
Both $\mu$ and $\chi$ are identified up to an unphysical arbitrary constant, $A(y) \to A(y) + a$ being a symmetry of the system.
We will fix it by requiring A(0)=0.
Besides, reparametrizations of the dilaton field should not change physical quantities, and when convenient we will simply take $\chi = e^{-ky_1}$ (see also Appendix \ref{App:kinetic}).

The background has to solve the bulk equations of motion 
\begin{eqnarray}
4 A'^2-A'' &=&-\frac{2\kappa^2}{3} V(\phi ) \nonumber \\
A'^2 &=& \frac{\kappa^2 \phi'^2}{12}-\frac{\kappa^2}{6} V(\phi ) \nonumber \\
\phi''&=& 4 A' \phi' +\frac{\partial V}{\partial \phi}.
\label{eq:bulkEOM}
\end{eqnarray}
The BC's (assumed to be $Z_2$-symmetric) are then:
\begin{eqnarray}
2 A'|_{y=y_0,y_1}&=& \pm \frac{\kappa^2}{3} V_1 (\phi )|_{y=y_0,y_1} \label{eq:BC1}\\
2\phi '|_{y=y_0,y_1} &=& \pm \frac{\partial V_1}{\partial \phi}|_{y=y_0,y_1}, \label{eq:BC2}
\end{eqnarray}
where the $+$ sign is for the UV brane and the $-$ sign for the IR brane. 

Let us now calculate the effective potential for the dilaton in these general backgrounds. The effective potential is obtained by integrating the bulk action over the solutions of the bulk equations of motion, with the scalar BC's (\ref{eq:BC1}) imposed both at the UV and the IR. We do not impose the Israel junction conditions (\ref{eq:BC1}) corresponding to the BC for the warp factor. Eventually the UV brane junction condition can be imposed thereby fixing the location $y_0$ of the UV brane, and possibly at the price of tuning the UV brane tension. 
The effective potential in terms of the general warp factor $A(y)$ and the general scalar background $\phi (y)$ is then given by 
\begin{equation}
V_{eff}(\chi ) =-2 \int_{y_0}^{y_1} dy \sqrt{g} \left[ -\frac{1}{2\kappa^2} (20 A'^2 -8 A'') -\frac{1}{2} \phi'^2 -V(\phi )\right] +\sqrt{g} V|_0 +\sqrt{g} V|_1
\end{equation}
Here we have integrated over the full circle rather than just over the orbifold. Special attention has to be paid to the singular pieces in $A''$ at the two boundaries, which will give an additional contribution to the effective potential of 
\begin{equation}
V_{eff}^{(sing)} = \left[ \sqrt{g} \frac{8 A'}{\kappa^2}\right]_0^1 
\end{equation}
while using the bulk equations of motion in (\ref{eq:bulkEOM}) the smooth part of the bulk is given by 
\begin{equation}
V_{bulk} = \frac{2}{\kappa^2} \int_{y_0}^{y_1} dy e^{-4A(y)} (4 A'^2-A'') = -\left[ \sqrt{g} \frac{2}{\kappa^2} A' \right]_0^1 \ .
\end{equation}
As expected, the entire effective potential is a boundary term, given in terms of the location of the IR brane $y_1$ by 
\begin{equation}
V_{eff}=V_{UV}+V_{IR}
\end{equation}
with 
\begin{equation}
V_{UV/IR}= e^{-4 A(y_{0,1})}  \left[ V_{0,1} \left( \phi(y_{0,1})\right) \mp\frac{6}{\kappa^2} A' (y_{0,1} )\right] \ .
\label{eq:effpotentialy}
\end{equation}
An alternative derivation of this effective potential using the Gibbons-Hawking boundary action is given in Appendix~\ref{App:GH}.
As expected, this potential vanishes for a solution that actually satisfies the boundary conditions (\ref{eq:BC1}) which we have not yet imposed. Once those are satisfied one has a flat solution to the bulk equations of motion and the resulting effective 4D cosmological constant necessarily vanishes.  This does not mean that the entire potential identically vanishes, nor does it imply that the minimum of the potential has to be at zero. 
In terms of the dilaton field $\chi= e^{-A(y_1)}$ and the location of the UV brane $\mu_0 = e^{-A(y_0)}$ (which effectively acts as UV cutoff regulator), the effective potential is
\begin{equation}
V_{IR} = \chi^4 \left[ V_1 \left( \phi \left( A^{-1} (-\log \chi)\right)\right)  +\frac{6}{\kappa^2} A'\left( A^{-1} (-\log \chi)\right)\right] \ .
\label{eq:effpotentialchi}
\end{equation}
while $V_{UV}$ is obtained by $\chi\to \mu_0$ and a sign flip in front of the $A'$ term.
The form of this potential is in accordance with the expectation that the general dilaton potential of a spontaneously broken conformal theory should be of the form~\cite{ourdilaton}
\begin{equation}
V_{eff}(\chi )= \chi^4 F(\lambda (\chi )),
\label{eq:effpotgeneric}
\end{equation}
where $\lambda$ is a coupling that introduces an explicit breaking of scale invariance.
Therefore we can make the holographic identification
\beq
F = V_1 + \frac{6}{\kappa^2} A' \ .
\label{eq:iden1}
\eeq

In the case of pure spontaneous breaking the potential should just be a pure quartic, which must  vanish if there is a stable vacuum in which scale invariance is spontaneously broken. For example in the case of pure AdS space without a scalar field (the original RS1 setup) the effective potential is indeed a pure quartic.  In this case, we have $A'=k$, and $V_1 (\phi ) =\Lambda_1$ (the IR brane potential is just a pure tension) and the effective dilaton potential is 
\begin{equation}
V_{dil, RS}= \chi^4 \left( \Lambda_1+\frac{6k}{\kappa^2} \right)\ .
\end{equation}
This pure quartic must vanish for the IR brane to not fly away or collide with the UV brane. From the 5D point of view the vanishing of this quartic is interpreted as the second fine tuning of RS. 

The minimization condition of the dilaton potential \eq{eq:effpotgeneric} can be written as
\beq
\left. \frac{d V_{eff} (\chi)}{d \chi} \right|_{\chi=\vev{\chi}} =  0 \ ,
\eeq
with
\beq
\frac{d V_{eff} (\chi)}{d \chi} = \chi^3\left[4 F +  \frac{\partial F}{\partial \lambda} \beta \right]\ , \quad \beta = \frac{\partial \lambda}{\partial \log \chi}
\eeq
Since we will require that the potential is minimized, we see that at the minimum
\begin{equation}
F = -\frac{1}{4} \frac{\partial F}{\partial \lambda} \beta
\end{equation}
implying that the potential at the minimum will be proportional to the value of the $\beta$-function. We will derive explicitly this same result from \eq{eq:effpotentialchi} in Section \ref{sec:CPR}.  That the value at the minimum itself might be non-vanishing implies that the solution does not actually have flat 4D sections, therefore to find the corresponding complete bulk solution a more general ansatz different from (\ref{eq:ansatz}) would be needed, along the lines of~\cite{Nemanja}.


\section{Constant bulk potential - flat dilaton potential by tuning two condensates\label{sec:nobulkmass}}
\setcounter{equation}{0}

Before we discuss the case with a non-trivial scalar bulk potential, it is very instructive to consider the theory with a constant potential. This is useful for two reasons:
\begin{itemize}
\item It provides a 5D gravity dual for the formation of a dimension four condensate and hence a ``soft-wall" version of the RS-model of SBSI.

\item This solution will be relevant for the IR region for the discussion of the general case with a small bulk mass in the next section.

\end{itemize}

The theory with constant bulk potential corresponds to adding an additional exactly marginal operator to the theory. If this operator condenses, it is expected to give another $\chi^4$ quartic term to the dilaton potential. For the case with a finite UV brane one also generically expects additional terms suppressed by the UV scale $\mu_0$. This will provide us with an alternative way of obtaining a flat dilaton potential compared to RS/GW. In GW one tunes the IR brane tension against the bulk cosmological constant to ensure that the condensate corresponding to the IR brane does not produce a quartic dilaton term, resulting in a flat dilaton potential. The other possibility considered here is to not impose the RS tuning at the IR brane, allowing a tree-level quartic from the condensate, but then canceling this with another quartic corresponding to the condensate of the bulk scalar. By appropriately tuning the the two condensates against each other one finds another way of obtaining a flat dilaton potential. While this also involves tuning, the significance of this is that by introducing the small bulk mass this tuning can be alleviated.

  We parametrize the bulk potential as 
\beq
V(\phi) = \Lambda_{(5)} = -\frac{6k^2}{\kappa^2} \ .
\eeq
For concreteness we will choose quadratic brane potentials, 
\beq
V_i(\phi) = \Lambda_i + \lambda_i (\phi - v_i)^2 \ ,
\label{eq:branepot}
\eeq
though for most arguments the detailed form of the brane potentials will not matter. The bulk only depends on the derivative of the scalar field, and thus one has a $\phi \to \phi +C$ shift symmetry, which signals the presence of conformal symmetry in this case. Thus one expects this to correspond to a purely spontaneous breaking of scale invariance.

The bulk equations of motion for this case can be solved analytically and the solutions are~\cite{CEGH}
\begin{eqnarray}
\label{eq:Aeps=0}
 A(y)&=& -\frac{1}{4} \log \left[ \frac{\sinh 4k (y_c-y)}{\sinh 4k y_c} \right] \\
\nonumber \\
\label{eq:phieps=0}
\phi(y) &=& -\frac{\sqrt{3}}{2\kappa} \log \tanh[2k (y_c-y)] +\phi_0 \ .
\end{eqnarray}
In this expression the (unphysical) constant in the warp factor was fixed such that $A(0)=0$. 
This solution describes the formation of a 4-dimensional condensate corresponding to the operator ${\cal O}$ that $\phi$ couples to. The singularity at $y_c$ corresponds to this condensate.  This solution on its own can be considered a ``soft-wall" version of a model of SBSI. While RS corresponds to the condensation of an infinite dimensional operator (hence the metric is exactly AdS all the way till the condensate forms, described by the appearance of the IR brane), here we have the more realistic case of the formation of a dimension four condensate. Both of these correspond to pure spontaneous breaking of scale invariance, and hence both of these should give pure quartic potentials for the dilaton. In our construction we will assume that both condensates are present, and that the pure RS condensate forms earlier, hence the IR brane will shield the singularity. Therefore we consider the region  $y<y_c$, and the location of the IR brane $y_1$ appears before the singularity, $y_1<y_c$: the RS  condensate in the CFT forms at a higher energy scale than the ${\cal O}$ condensate.  

For finite $y_c$, the AdS boundary is at $y=- \infty$, 
\beq
A'(y \to - \infty)=k \ , \quad  \phi(y \to - \infty) = \phi_0 \ .
\eeq
Exact AdS space is only obtained in the limit $y_c \to \infty$,
\beq
\lim_{y_c \to \infty} A'(y) = k  \ , \quad \lim_{y_c \to \infty} \phi(y) = \phi_0 \ .
\label{eq:AdS}
\eeq
The scalar profile is constant in this limit. The AdS limit \eq{eq:AdS} can only be obtained by imposing that both brane potentials are pure tensions (no $\phi$-dependence) and the tensions obey the RS tunings:
\beq
V_i(\phi) = \mp \frac{\Lambda_{(5)}}{k}\ ,
\label{eq:RStuning}
\eeq
in which case  the singularity is pushed to $y_c\to \infty$.

For generic brane potentials $y_c$ will be finite, thus the space will deviate from pure AdS. We want to find the effective potential for the dilaton field in this case.
A convenient parameterization of the the dilaton $\chi$ and the location of the UV brane $\mu_0$   is
\begin{equation}
\chi^4 = e^{-4 A(y_1)} = \frac{\sinh 4k(y_c-y_1)}{\sinh 4ky_c}, \ \ 
\mu_0^4 =e^{-4 A(y_0)}=  \frac{\sinh 4k(y_c-y_0)}{\sinh 4ky_c} \ ,
\label{eq:mu0def}
\end{equation}
while for the location of the singularity we will use the parametrization
\begin{equation}
\delta^4 = \frac{1}{\sinh 4ky_c}.
\label{eq:deltadef}
\end{equation}
To determine the effective potential we need to impose the BC's for the scalar field \eq{eq:BC2}. For concreteness we can choose simple quadratic brane potentials \eq{eq:branepot}, 
though the specific form of the brane potentials will not be important. For these potentials the scalar boundary conditions are 
\begin{equation}
2 \lambda_i \left(\phi_0 -\frac{\sqrt{3}}{2\kappa} \log\tanh [2k(y_c-y_i)] -v_i\right)= \mp \frac{2\sqrt{3}k}{\kappa} \frac{1}{\sinh 4k(y_c-y_i)}
\end{equation}
These should be used to determine the constants $y_c$ and $\phi_0$ for use in the effective potential. Since both of these equations depend only on the distances of the brane to the singularity $y_i-y_c$ both of them can be written in terms of the combination of the variables $\chi^4/\delta^4$ and $\mu_0^4/\delta^4$. We can use the UV scalar equation to determine $\phi_0$ in terms of the location of the UV brane as 
\begin{equation}
\phi_0 = v_0 \left( 1+ f_0 (\delta^4/\mu_0^4)\right)
\end{equation}
since in the simultaneous limit $\delta \to 0$ and $\mu_0 \to \infty$, $\phi_0$ approaches $v_0$. The IR brane equation can then be used to separately determine $\delta$, and the result will be of the form
\begin{equation}
\delta^4 = \chi^4  f_1 (\phi_0, v_1,\lambda_1).
\end{equation}
Combining these two equations we find that the structure of the solutions to the scalar BC's will be of the form
\begin{eqnarray}
\phi_0 &=& v_0 \left( 1 + {\cal O}( \chi^4/\mu_0^4 ) \right), \label{eq:phi0} \\
\delta^4 &=& \chi^4 f_1 \left(  v_0 ( 1 + {\cal O}( \chi^4/\mu_0^4 ) ), \lambda_1, v_1\right). \label{eq:delta}
\end{eqnarray}
These expressions have the right limits to be identified with an external source and a condensate:
\begin{eqnarray}
&& \lim_{\mu_0 \to \infty} \phi_0 = v_0  \ ,  \\
&& \lim_{\chi \to 0} \delta^4 = 0\ .
\end{eqnarray}

For example in the limit $\lambda_{0,1}\to \infty$ we find 
\begin{equation}
\phi_0=v_0 +\frac{\sqrt{3}}{2\kappa} \log \left( \sqrt{1+\frac{\delta^8}{\mu_0^8}}-\frac{\delta^4}{\mu_0^4} \right), \ \ 
\delta^4 =\chi^4 \sinh \left( \frac{2\kappa}{\sqrt{3}}(v_1-\phi_0) \right) \ ,
\label{eq:deltalambda}
\end{equation}
and the system can be exactly solved, although the exact expressions are not important for the general argument. 

The full effective dilaton potential is
\beq
V_{eff} = V_{UV} + V_{IR}
\eeq
with
\beq
V_{UV}=\mu_0^4 \left[ \Lambda_0 -\frac{6k}{\kappa^2} \sqrt{1+\frac{\delta^8}{\mu_0^8}} +
\lambda_0 \left( \phi_0-v_0 
-\frac{\sqrt{3}}{2\kappa}\log \left[ \sqrt{1+\frac{\delta^8}{\mu_0^8}}-\frac{\delta^4}{\mu_0^4}\right]
\right)^2 \right]
\label{VUV}
\eeq
\beq
V_{IR}=\chi^4 \left[ \Lambda_1 +\frac{6k}{\kappa^2} \sqrt{1+\frac{\delta^8}{\chi^8}} +
\lambda_1 \left( \phi_0-v_1 
-\frac{\sqrt{3}}{2\kappa}\log \left[ \sqrt{1+\frac{\delta^8}{\chi^8}}-\frac{\delta^4}{\chi^4}\right]
\right)^2 \right].
\label{VIR}
\eeq\\

We can see that using (\ref{eq:delta}) the IR term will become a pure quartic modulo the $\chi$-dependence of $\phi_0$ that is suppressed by the location of the UV brane, while the UV contribution will be a pure cosmological constant given by the RS tuning, and additional $\chi^4/\mu_0^4$-type corrections:
\begin{eqnarray}
V_{IR} &=& \chi^4 \left( a(v_0) +{\cal O}( \chi^4/\mu_0^4) \right) \label{eq:VIR}\\
V_{UV} &=& \mu_0^4 \left(\Delta_0 + {\cal O}( \chi^8/\mu_0^8) \right), \label{eq:VUV} 
\end{eqnarray}
where $a (v_0)$ is a constant that determines the quartic dilaton coupling, which depends on the UV value of the scalar field $v_0$ (and all the other parameters of the theory), while $\Delta_0$ is the usual RS UV fine tuning condition $\Delta_0  = \Lambda_0-6k/\kappa^2$. For generic values of the parameters this potential would be minimized for $\chi \sim \order{\mu_0}$ and thus no hierarchy would be generated. 

Again for the sake of illustration, in the limit $\lambda_{0,1}\to \infty$ one finds the potentials
\beq
V_{UV}= \mu_0^4 \left[ \Lambda_0 - \frac{6k}{\kappa^2} \left(1+ \frac{\chi^8 \sinh^2 \left( \frac{2\kappa}{\sqrt{3}}(v_1-\phi_0) \right) }{\mu_0^8+\chi^8-2\mu^4\chi^4\cosh \left( \frac{2\kappa}{\sqrt{3}}(v_1-\phi_0) \right)} \right)^{1/2} \right] \ ,
\eeq
\beq
V_{IR}= -V_{UV} (\mu_0 \leftrightarrow \chi, \Lambda_0 \to -\Lambda_1),
\eeq
and therefore the quartic dilaton coupling reads
\beq
a(v_0)= \Lambda_1 + \frac{6k}{\kappa^2} \cosh \left( \frac{2\kappa}{\sqrt{3}}(v_1-v_0) \right) \ .
\eeq
This can be made to vanish by properly tuning the UV value of the scalar, $v_0$, which is the holographic equivalent to a tuning of the initial value of the external perturbation, $\lambda(\mu_0) \mathcal{O}$. It is particularly illuminating to notice that in the limit $\lambda_1 \to \infty$ we have taken, the whole IR potential comes from the $(6/\kappa^2)A'$ piece, that is from the back-reaction on the metric. This is easy to understand since the IR $\phi$ BC fixes $\phi' \sim \partial V_1/\partial \phi$ and due to the structure of $V_1$ one has $V_1 \sim \phi'^2/\lambda_1 \to 0$ when $\lambda_1 \to \infty$.

The generic structure of the effective potential has a very clear explanation: the only explicit breaking of scale invariance in this theory corresponds to the introduction of the UV brane.  Thus in the limit when the UV brane is removed, the effective potential must reduce to a pure quartic (plus a UV contribution to the cosmological constant). This is indeed what we find here, and the explicit expression for the quartic depends on $v_0$, the value of the scalar field in the UV. One can make the entire potential vanish by tuning the UV cosmological constant to zero, and by tuning $v_0$ appropriately. The important difference in this tuning compared to Goldberger-Wise is that here we tune the UV value of the scalar field (that is the value of the perturbing coupling in the UV), rather than the IR brane tension (which is arbitrary here). We will see in the next section that this tuning will be alleviated once we let the perturbing coupling run, that is once we include a non-trivial potential for $\phi$, in particular a mass term, $m^2 \sim \epsilon k^2$.
Then $v_0 \to v_0 (\chi/\mu_0)^\epsilon$, which will become the leading order term in $\chi/\mu_0$, and will then set the hierarchy. 

We should stress that once the tuning on $v_0$ is imposed corresponding to setting the quartic to zero, $a(v_0)=0$, the spacetime (\ref{eq:ansatz}) with the warp factor given by  (\ref{eq:Aeps=0}), still represents the 5D dual of a spontaneously broken CFT, even though the metric deviates significantly from AdS:
\begin{equation}
ds^2= \sqrt{ \frac{\sinh 4k(y_c-y)}{\sinh 4k y_c}} dx^2 -dy^2 \ .
\end{equation}
That this metric corresponds to a spontaneously broken scale invariant theory should be clear from the previous analysis and the resulting effective potential for the dilaton, but one  can also  explicitly consider the effect of the scale transformation $y\rightarrow y+a$, $x\rightarrow e^\alpha(a) x$. If the IR brane is kept fixed, then this transformation will not leave the metric invariant simply due to the presence of the IR brane\footnote{The UV brane is a source of explicit breaking, which is eliminated once the UV brane is removed, $\mu_0\rightarrow\infty$. } - this is exactly what one expects from a spontaneous breaking of scale invariance. The symmetry is restored by simultaneously  moving the IR brane, $y_1\rightarrow y_1+a$. Due to the scalar BCs that result in (\ref{eq:delta}) a shift in $y_1$ should also be accompanied by a shift in $y_c$, which will make the shift in the warp factor $y$-independent:  the net shift in the warp factor is then compensated  by the scale factor $e^\alpha(a)=\left[\sinh(4k y_c)/\sinh(4k(y_c+a))\right]^{1/2}$. 
\footnote{The reader may notice that $e^\alpha(a)$ is mildly dependent on $y_1$ so that the scale transformation of the dilaton field is slightly non-linear, $\chi\rightarrow f(\chi)\chi$,  with $f(\chi)$ a slowly varying function. One might then argue that a more natural  parametrization of the dilaton field is provided by $\chi=\mathrm{Exp}[-k y_1]$ which transforms covariantly  even though it  does not seem to reproducing the expected quartic potential. In fact, in  App.~\ref{App:kinetic}, we clarify these points and show how both parametrizations are legitimate and give rise to a purely quartic potential once the kinetic mixing with gravity is properly taken into account.}

Notice that in order to obtain a small cosmological constant (neglecting $\order{\chi^8/\mu_0^4}$ terms), we have to impose the UV RS tuning $\Delta_0 \ll 1$. 
This condition is actually also needed in order to obtain a suitable dilaton potential, due
to the presence of a dilaton-gravity kinetic mixing, of $\order{\chi^2/\mu_0^2}$ (see Appendix~\ref{App:kinetic}). If the UV RS tuning is not imposed we generate a term $\Delta_0 \mu_0^2 \chi^2$ in the potential, which would not allow for the generation of a large hierarchy between $\mu_0$ and $\chi$.

In two appendices,~\ref{App:smallback} and~\ref{App:GW}, we present the detailed description of the cases with a small back-reaction and no bulk mass, and small back-reaction and small bulk mass (the GW case). 

\section{Light dilaton without tuning: the general case\label{sec:CPR}}
\setcounter{equation}{0}

We are now ready to consider the general case with $\order{1}$ IR brane mistuning, a large condensate and long slow running of the scalar due to a small scalar bulk mass. The bulk scalar potential is again given by 
\beq
V(\phi) = -\frac{6k^2}{\kappa^2} - 2 \epsilon k^2 \phi^2 \ .
\label{eq:bulkpot}
\eeq
We want to stress again that the exact form of the perturbing bulk potential does not matter, as long as it is always parametrically suppressed (that is $\epsilon$ multiplies the entire bulk potential). For more complicated potentials the form of the RGE running will change, but as long as the $\epsilon$ suppression persists the running will be mild. CPR suggested that the overall suppression of the bulk potential by $\epsilon$ may be due to $\phi$ being a 5D bulk Goldstone field and $\epsilon$ is the parameter of a small explicit breaking term. 

For the  brane potentials we will again use a quadratic expression, \eq{eq:branepot}, but as explained before the detailed form of this potential again does not matter.

In order to find the bulk solution, we note that we can break up the bulk into two regions: the UV region dominated by a mild RGE running of the scalar where the solution remains close to AdS, and the IR region dominated by the condensate, where the solution is of the form considered in the previous section. We will then match up these two solutions using asymptotic matching for the boundary layer theory of differential equations~\cite{Bender}. 

The UV solution is characterized by a mild running of the scalar, which means that one can neglect the second derivative of the scalar: $\phi', \delta V(\phi) \gg \phi''$. The deviation from AdS space is small, so in this region $A' =k$, and the scalar equation is first order:
\begin{equation}
k \phi' - \epsilon \phi =0
\end{equation}
so the solution in the UV region (which we call the ``running region" and denote by subscript $r$) is given by
\begin{eqnarray}
A'_r(y) &=& k \\
\phi_r(y)  &=& \phi_0 e^{\epsilon k y}.
\end{eqnarray}
This solution is self-consistent in the UV as long as the back-reaction on the metric is negligible, that is $\kappa^2 \epsilon k^2 \phi^2 /3 \ll A'^2$, which restricts the region of validity to 
\beq
 y \ll \frac{1}{\epsilon  k} \log \left( \frac{1}{\sqrt{\epsilon}\phi_0 \kappa}\right) \ .
 \eeq
The second region where we can find an analytic solution is the region where the condensate dominates. In this case the behavior of the scalar is dominated by the $\phi'',\phi'$ terms and the additional bulk potential is negligible. In this case we recover the equations for the zero bulk mass considered in the previous section. Thus there is a universality in the IR behavior of the solution, since it is dominated by the dimension 4 condensate. Therefore in this IR ``condensate region" (denoted by the subscript $c$) the solution is given by
\begin{eqnarray}
A'_c(y) &=& -k \coth \left( 4k(y-y_c) \right) \\
\phi_c(y)  &=& \phi_m - \frac{\sqrt{3}}{2 \kappa} \log \left( -\tanh \left( 2k(y-y_c) \right) \right),
\end{eqnarray}
where $\phi_m$ is the matching value of the scalar field. Applying the method of asymptotic matching for a boundary layer theory we obtain the matching conditions:
 \begin{eqnarray}
\lim_{y \to -\infty} \phi_c = \lim_{y \to y_1} \phi_r \quad  &\Rightarrow& \quad  \phi_m = \phi_0 e^{\epsilon k y_1}\\
\lim_{y \to -\infty} A'_c = \lim_{y \to y_1} A'_r  \quad  &\Rightarrow& \quad k=k
\end{eqnarray}
The details of this matching are explained in Appendix~\ref{App:Matching}.

As before, to determine the constants $\phi_0$ and $y_c$ we impose the  UV BC for $\phi_r$ and the IR BC for $\phi_c$:
\begin{eqnarray}
2\phi'_r |_{y=y_0} &=& + \frac{\partial V_{0}}{\partial \phi}|_{\phi(y)=\phi_r(y_0)} \ , \\
2\phi'_c|_{y=y_1} &=& - \frac{\partial V_{1}}{\partial \phi}|_{\phi(y)=\phi_c(y_1)} \, 
\label{eq:phiBCs}
\end{eqnarray}
from which we find, in the limit $\lambda_0, \lambda_1 \to \infty$,
\begin{eqnarray}
\phi_0 &=& v_0 \mu_0^{\epsilon} \ , \\
\delta &=& \chi \tanh^{1/4} \left( \frac{\kappa}{\sqrt{3}}(v_1-\phi_m) \right) .
\label{eq:solphiBCs}
\end{eqnarray}
To simplify our expressions we have used the alternate definition of the dilaton, the UV scale and the condensate $\mu_0 = e^{-ky_0}$, $\delta = e^{-ky_c}$, and $\chi = e^{-ky_1}$. As we learned from the constant bulk potential case, the distance between the singularity and the IR brane, or equivalently $\delta/\chi$, depends on the IR potential parameters, in particular on the difference between $\phi(y_1) = v_1$, and $\phi(y_0) = v_0$, where the latter is now modulated by $(\mu_0/\chi)^{\epsilon}$.\\

\begin{figure}[!t]
\begin{center}
\includegraphics[width=3.0in]{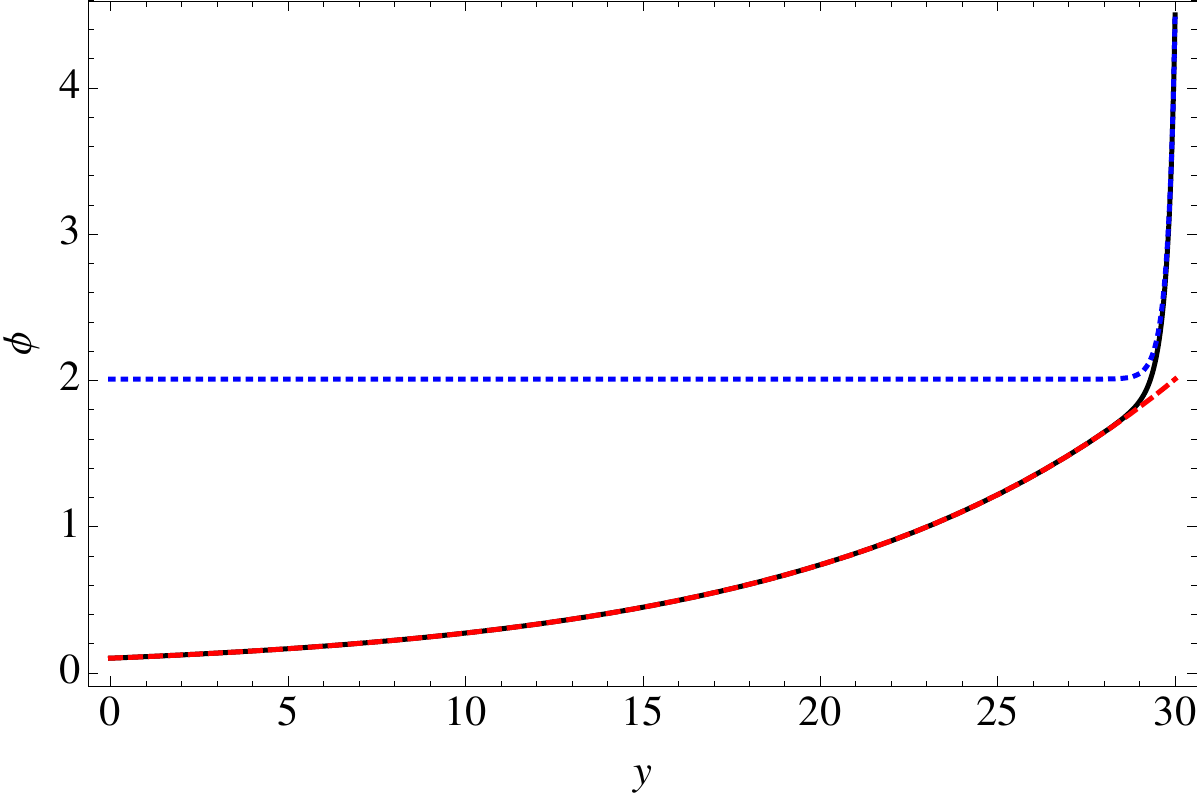}
\includegraphics[width=3.0in]{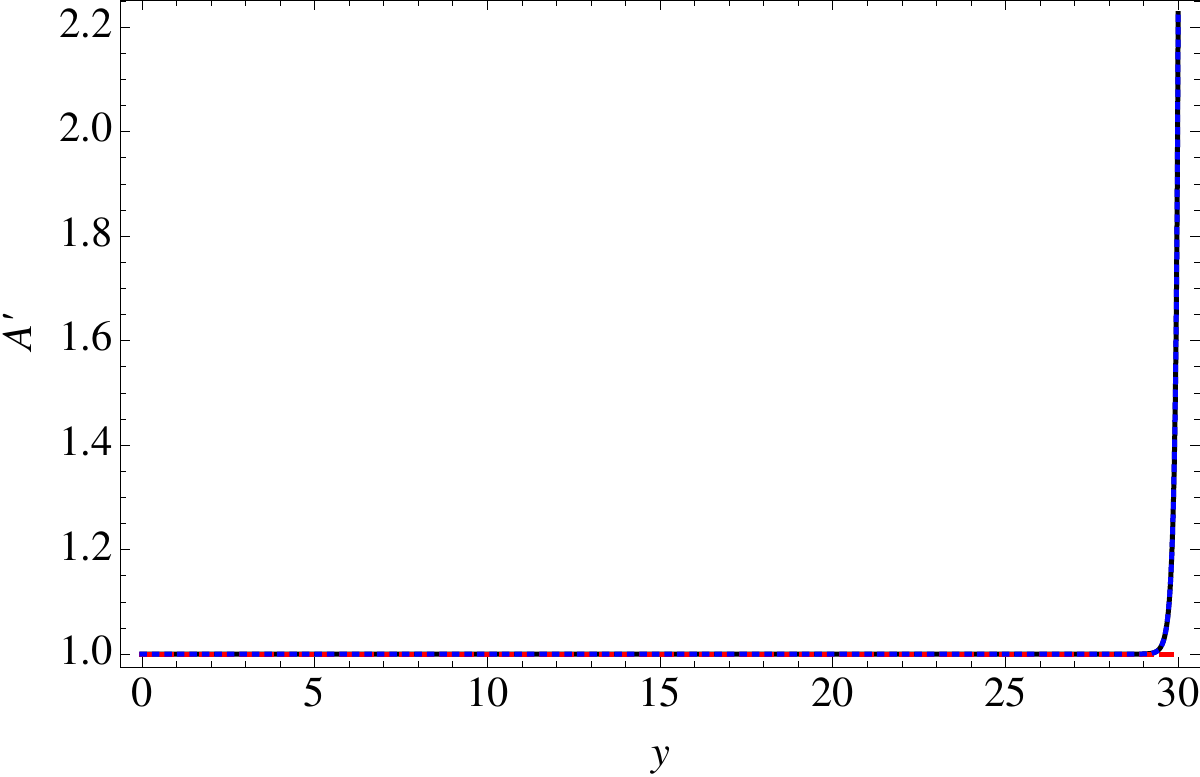}
\caption{Left, bulk scalar profile: $\phi_{full}$ (solid black), $\phi_{r}$ (dashed red), and $\phi_b$ (dotted blue).
Right, effective AdS curvature, $A'(y)$: same color code.}
\label{phi}
\end{center}
\end{figure}

The full approximate solution\footnote{We have  dropped a term $\frac{\sqrt{3}}{2 \kappa} \log \left( \tanh \left( 2k(y_c-y_0)\right)\right)$ which is exponentially small for $y_c \gg y_0$, but which strictly ensures $\phi(y_0)=v_0$. This term would be automatically included if the matching of the $\phi_r$ was at $y=y_0$ instead of $y \to -\infty$. This approximation propagates to \eq{eq:solphiBCs}, and amounts to unimportant $\order{\chi/\mu_0}$ corrections.} to the system is 
\beq
\phi_{full}(y) &=& \phi_r(y) + \phi_c(y) - \phi_m \\
&=&v_0 \,e^{\epsilon k (y-y_0)} - \frac{\sqrt{3}}{2 \kappa} \log \left( \tanh \left( 2k(y_c-y) \right) \right)
\label{eq:Phifull}
\eeq
 and equivalently for $A'(y)$.
In $z=e^{-ky}$ coordinates these are
\begin{eqnarray}
A'_{full}(z)&=& \left(-1 + \frac{2 z^8}{z^8 + \chi^8 \tanh^2 \left( \frac{\kappa}{\sqrt{3}}(v_1-v_0 (\mu_0/\chi)^{\epsilon}) \right)
} \right)^{-1} \ ,
 \\
\phi_{full}(z)&=& v_0 \left( \frac{\mu_0}{z} \right)^{\epsilon} -  \frac{\sqrt{3}}{2 \kappa} \log \left[
 -1 + \frac{2 z^4}{z^4 + \chi^4 \tanh \left( \frac{\kappa}{\sqrt{3}}(v_1-v_0 (\mu_0/\chi)^{\epsilon})\right)} 
 \right] \ .
\end{eqnarray}
This solution exhibits the correct asymptotic behavior. 
We can see this explicitly in Fig.~\ref{phi}. The full solution interpolates nicely between the running and the condensate dominated solutions.
\\

We can now compute the effective potential for the dilaton as usual (again in the $\lambda_{0,1}\to \infty$ limit)
\begin{eqnarray}
V_{UV} &=& \mu_0^4 \left[ \Lambda_0
-\frac{6k}{\kappa^2}
\right] \ , 
\label{eq:UVpot} \\
V_{IR} &=& \chi^4 \left[ 
\Lambda_1 + \frac{6 k}{\kappa^2} \cosh \left( \frac{2 \kappa}{\sqrt{3}} (v_1 - v_0 (\mu_0/\chi)^{\epsilon}) \right)
 \right] \mathrm{sech}^2 \left( \frac{\kappa}{\sqrt{3}} (v_1 - v_0 (\mu_0/\chi)^{\epsilon} )\right). \ \ \ \ \ \ 
\label{eq:IRpot}
\end{eqnarray}
The UV effective potential contains a constant piece, which must be tuned to zero in order to obtain a flat 4D space (usual UV RS tuning).
The IR potential is of the expected form $\chi^4 F[(\mu_0/\chi)^{\epsilon}]$.
This is the leading part of the potential, whose minimization will determine the position of the minimum, $\vev{\chi}$, up to $\order{\epsilon}$ corrections.
Recall also that the potentials \eq{eq:UVpot} and \eq{eq:IRpot} are corrected by $\order{\chi^2/\mu_0^2}$ once the dilaton-gravity kinetic mixing is fully included, see Appendix~\ref{App:kinetic}.
It is therefore important to tune $\Lambda_0 \simeq 6 k / \kappa^2$ in order not to generate a large $\chi^2$ term.

\begin{figure}[!t]
\begin{center}
\includegraphics[width=2.6in]{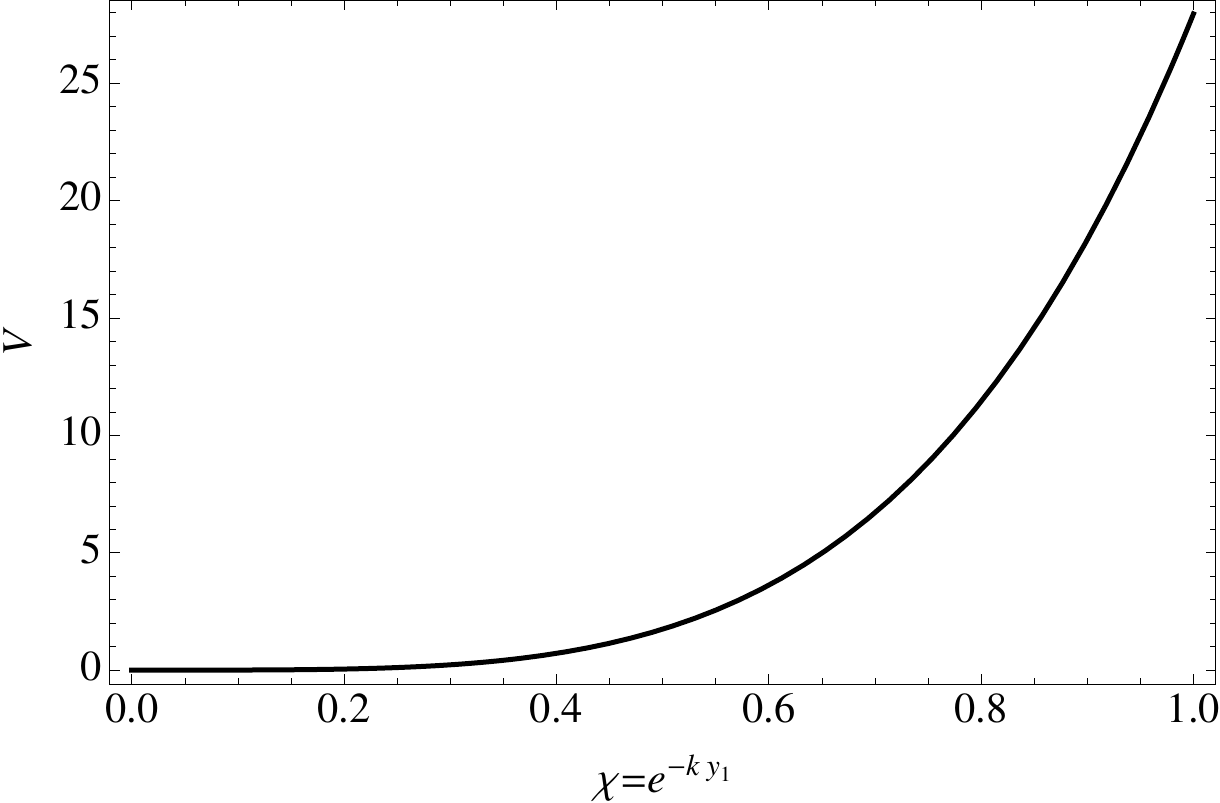}
\hspace{0.3cm}
\includegraphics[width=3.05in]{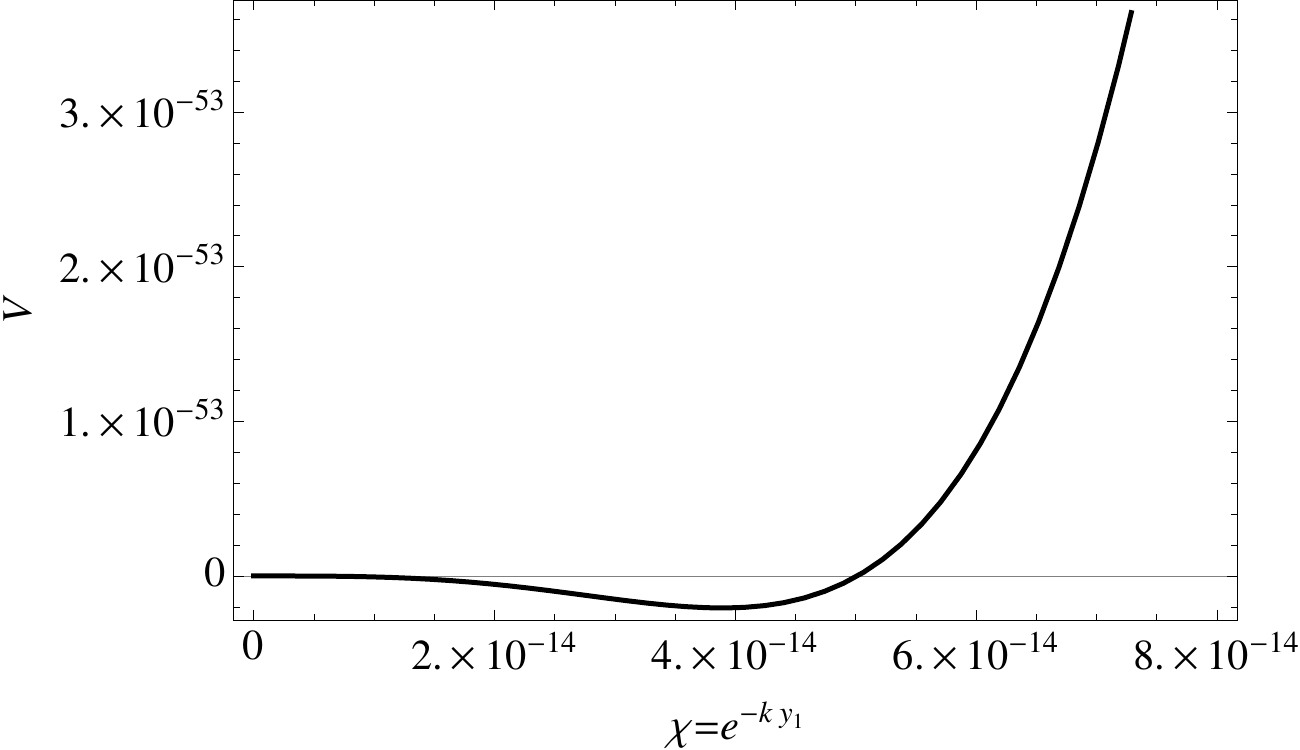}
\caption{The plot of the effective dilaton potential \eq{eq:IRpot} for the parameters $\epsilon=0.1$, $v_0=0.1$, $v_1=4.5$, $\Lambda_1 = - 50$, $\mu_0 = 1$, and $\kappa = 0.5$, all of them in units $k=1$. The plot in the right is a zoom of the region where the minimum of the potential is. \label{fig:potential}}
\end{center}
\end{figure}

To leading order in $\epsilon$, the condition for the minimum of the potential is 
\beq
\frac{\partial V_{IR}}{\partial \chi} = \chi^3 \left( 4 F[(\mu_0/\chi)^\epsilon] + F'[(\mu_0/\chi)^\epsilon] \epsilon (\mu_0/\chi)^{\epsilon} \right) = 0 
 \label{eq:minimization}
\eeq
leading to a dilaton VEV
\beq
\frac{\vev{\chi}}{\mu_0} = \left( 
\frac{v_0}{v_1 - \, \mathrm{sign}(\epsilon) \frac{\sqrt{3}}{2 \kappa} \, \mathrm{arcsech} (-6k/\kappa^2 \Lambda_1)}
\right)^{1/\epsilon} + \order{\epsilon}
\label{eq:minimum}
\eeq
while the potential will be obviously of order $F[(\mu_0/\chi)^\epsilon] = {\cal O}(\epsilon)$.
Notice that for this to be a good minimum we need $\Lambda_1 < 0$ and $|\Lambda_1| > 6k/\kappa^2$. 
One can clearly see from \eq{eq:IRpot} that if these conditions are not satisfied then the effective quartic is always positive $F[\chi/\mu_0]> 0$ for all $\chi$, and the minima can only be found at $\vev{\chi}=0$ or $\vev{\chi}=\mu_0$.
Furthermore, in order for the effective quartic to be positive at $\chi = \mu_0$ (thus avoiding this as a minimum), one must have $|\Lambda_1| < \frac{6k}{\kappa^2} \cosh ( \frac{2 \kappa}{\sqrt{3}} (v_1 - v_0) )$.
This condition is easily satisfied, either if $v_1 \gg v_0$, a condition consistent with $\epsilon > 0$, or $v_0 \gg v_1$, consistent with $\epsilon < 0$.
However, notice that a large hierarchy, which in this scenario it is given by the point where $6A'/\kappa^2$ compensates $\Lambda_1$, is easier to produce for the case $\epsilon > 0$, since in this case $v_1 - v_0(\mu_0/\chi)^\epsilon$ runs slower than for $\epsilon<0$.  This is the scenario we have advocated for naturally canceling a large quartic at the scale $\mu_0$. We show a plot of the potential (\ref{eq:IRpot}) in Fig.~\ref{fig:potential}, where we can see that a shallow stable minimum with a small mass is indeed generated.

The dual CFT interpretation of the potential \eq{eq:IRpot} for the interesting $\epsilon > 0$ is simple. 
The quartic in the absence of perturbation (that is $v_0 = 0$) is given by $F_0 = \Lambda_1 + \frac{6k}{\kappa^2} \cosh (\frac{2 \kappa}{\sqrt{3}}v_1)$.
This is generically large and positive, hence there is no SBSI at high scales.
Once the perturbation is turned on, it grows larger in the IR, $v_0 (\mu_0/\chi)^{\epsilon}$.
This in turn decreases the effective quartic, until the minimum $F[\chi/\mu_0]={\mathcal O}(\epsilon)$ is found.
Effectively, the dilaton quartic coupling relaxes to zero at $\chi/\mu_0 \ll 1$. At this point SBSI will occur.

The dilaton mass, to leading order in $\epsilon$ is given by
 \beq
m_\chi^2 \sim \epsilon \frac{32 \sqrt{3} k v_0}{\kappa} \tanh \left( \frac{\kappa}{\sqrt{3}}(v_1 - v_0(\mu_0/\chi)^{\epsilon}) \right) \vev{\chi}^2 (\mu_0/\chi)^{\epsilon} + \order{\epsilon^2}
\label{eq:dilmass}
\eeq
One then concludes, that regardless of the size of the back-reaction on the metric, the dilaton remains light as long as the $\beta$-function is small.
Of course the actual physical mass of the dilaton also depends on the normalization of its kinetic term, which we have not calculated in this paper, we assume it is $\order{1}$ or bigger. The kinetic term normalization does not remove the $\epsilon$ suppression in (\ref{eq:dilmass}). 

Next we examine the value of the potential at the minimum, which is the effective cosmological constant. In the approximation we have followed in this section, the cosmological constant is given by $\Lambda_{eff}=V_{IR}(\vev{\chi})$ from \eq{eq:IRpot}, since we have fine-tuned away $V_{UV}$. The value of the IR potential at the minimum is
\begin{equation}
V_{IR}^{min} = 
- \epsilon \frac{2 \sqrt{3} k v_0}{\kappa} \tanh \left( \frac{\kappa}{\sqrt{3}}(v_1 - v_0(\mu_0/\chi)^{\epsilon}) \right) \vev{\chi}^4 (\mu_0/\chi)^{\epsilon}
\sim -m_{\chi}^2 \frac{\vev{\chi}^2}{16}
\end{equation}
As expected, the value of the minimum is suppressed by $\epsilon$, and also by $4-\epsilon$ powers of the dilaton at the minimum. Assuming that this is the origin of the hierarchy, that is $\langle \chi \rangle k \sim$ TeV,  the resulting potential is of order $\epsilon $ TeV$^4$. 
Therefore, since we have minimized the potential at $\order{\epsilon^0}$ \eq{eq:minimization}, then, $V_{IR}(\vev{\chi}) = \order{\epsilon \vev{\chi}^4 k^4}$.
Phenomenologically, this contribution is still too large unless $\epsilon \sim 10^{-60}$. Also, since $\epsilon > 0$ for most interesting applications, the IR potential is usually negative. Of course since our full potential contained a tuned value of the UV contribution,  
one  could have tuned $V_{UV} = \order{\epsilon \vev{\chi}^4 k^4}$ previously, such that eventually $\Lambda_{eff} = 0$ or small positive.
This change in the UV potential affects the minimization only at $\order{\epsilon}$, and thus it does not affect our conclusions.

We finally show that regardless of the explicit form of the IR brane potential the value of the potential at the minimum is always suppressed by $\epsilon$. The form of the dilaton potential is $e^{-4A(y_1)} F(y_1,y_c)$, hence the derivative of the potential is given by
\beq
\frac{\partial V_{IR}}{\partial y}|_{y_1} =e^{-4 A(y)} \left( -4 A'(y) F(y,y_c) + \frac{d}{dy}  F(y,y_c) +\frac{d}{dy_c}  F(y,y_c)  \frac{dy_c}{dy_1}   \right)|_{y_1} = 0 
\label{eq:minimizationy}
\eeq
Note that
\beq
\frac{d}{dy}  F(y,y_c)=
\frac{\partial V_1} {\partial \phi} \phi'+\frac{6}{\kappa^2} A'' ~,
\label{eq:miny}
\eeq
by using  the bulk equation of motion $A'' = \kappa^2 \phi'^2/3$, can be brought to a form proportional to  the scalar boundary condition (\ref{eq:BC2}), and thus vanishes at the IR brane.
Note also that the functional dependence of $F$ on $y_c$ comes in the form $y_c -y$, so in the $\epsilon \rightarrow 0$ limit we also have $\frac{d}{dy_c}  F(y,y_c)=0$.  Thus at the minimum
\beq
\frac{d}{dy_c}  F(y,y_c) = -\frac{d}{dy}  F(y,y_c)
+\epsilon\,  k \, \phi_0 \,e^{-\epsilon k (y-y_0) }\frac{\partial V_1} {\partial \phi} =\epsilon\,  k \, \phi_0 \,e^{-\epsilon k (y-y_0) }\frac{\partial V_1} {\partial \phi} 
\eeq
thus for the value of the potential at the minimum we find
\beq
F|_{min}= \frac{ \epsilon\,  k \, \phi_0 \,e^{-\epsilon k (y_1-y_0) }}{A'(y_1)} \frac{\partial V_1} {\partial \phi} 
 \frac{dy_c}{dy_1}  \ .
 \eeq

\section{Conclusions}

We presented a 5D holographic construction of a theory with a naturally light dilaton: a conformal theory perturbed by an almost marginal (dimension $4-\epsilon$) operator. As the coupling of the perturbation slowly increases through renormalization group running, the effective quartic of the dilaton slowly decreases. Around the scale where the effective quartic vanishes scale invariance will be broken, the perturbing operator (along with other CFT operators) will develop a condensate, and a stable minimum of the dilaton potential at hierarchically small scales. If the perturbing operator remains close to marginal even for large couplings, the dilaton mass squared and the value of the dilaton potential at the minimum will both be suppressed by $\epsilon$. 

In order to find the explicit holographic description of this setup we first considered the case with an exactly marginal perturbation, and described the exact solutions of the scalar-gravity equations for this system. This solution is a novel holographic dual of an exactly conformal theory where conformality is broken via the condensate of a dimension 4 operator. Even though the metric deviates significantly from AdS in the IR, this nevertheless corresponds to a non-linearly realized conformal theory. This solution provides the description of the IR region for the case with the $4-\epsilon$ dimensional condensate, while the UV is dominated by the slow running of the bulk scalar. Matching these solutions one obtains the full background for the system with the light dilaton. Finally we applied the formula for the effective dilaton potential derived earlier in this paper to verify that the dilaton mass squared and the contribution to the cosmological constant are both indeed suppressed by $\epsilon$. Phenomenologically $\epsilon$ cannot be taken small enough
to solve the cosmological constant problem since the dilaton must be heavier than about an eV, but this mechanism can improve on the SUSY suppression of the cosmological constant by many orders of magnitude.

\section*{Acknowledgements}
B.B. thanks Paolo Lodone  for discussions. 
B.B. is supported in part by the ERC Advanced Grant No.~267985, ``Electroweak Symmetry Breaking, Flavour and Dark Matter: One Solution for Three Mysteries'' (DaMeSyFla), the NSF Grant PHY11-25915, and by the MIUR-FIRB grant RBFR12H1MW. B.B. thanks the KITP at UCSB for the hospitality during the workshop ``Exploring TeV Scale New Physics with LHC Data'' where part of this work was completed.
C.C. and J.S. are supported in part by the NSF grant PHY-0757868.  B.B., J.H. and J.T. thank Cornell University for hospitality during the course of this work.  J.H. is supported in part by the DOE grant DE-FG02-85ER40237.  J.T. was supported by the
DOE grant DE-FG02-91ER406746.

\appendix
\section{Effective potential and boundary terms\label{App:GH}} 
\setcounter{equation}{0}

We show here another way to get the effective dilaton potential (\ref{eq:effpotentialy}) from integrating out the extra dimension.
In order to properly disentangle  brane and bulk contributions to $V_{eff}(\chi)$ it is convenient to explicitly write the Gibbons-Hawking boundary terms
\begin{equation}
S = S_{bulk}-\int d^4x \sqrt{g_0} V_0 (\phi) - \int d^4x \sqrt{g_1} V_1 (\phi)-\frac{1}{\kappa^2}\int d^4x\sqrt{g_0} K_0
-\frac{1}{\kappa^2}\int d^4x\sqrt{g_1} K_1
\end{equation}
where $K_{0,1}$ are the extrinsic UV and IR curvatures that  for a rigid brane $y=y_{0,1}$ in our warped metric are $K_{0,1}=\nabla_M n^M=\mp 4A^\prime(y_{0,1})$. $n^M=(0,\pm)$ is the normal unit vector and $-$(+) is for the UV (IR) brane. This contribution gets actually doubled, because of the orbifolding, after integrating out the extra dimension
\begin{equation}
V^{boundary}_{UV/IR}=e^{-4A(y_{0,1})}\left[V_{0,1}(\phi)\mp \frac{8}{\kappa^2} A^{\prime}(y_{0,1})\right]\,,
\end{equation}
and it adds to the bulk contribution $V_{bulk}=\pm 2/\kappa^2 A^\prime(y_{0,1})$ (after using the $\phi$ equations of motion)
\begin{equation}
V_{UV/IR}=e^{-4A(y_{0,1})}\left[V_{0,1}(\phi)\mp \frac{6}{\kappa^2} A^{\prime}(y_{0,1})\right]\,,
\end{equation}
giving the effective potential  (\ref{eq:effpotentialy}).

\section{The massless case for small back-reaction\label{App:smallback}}
\setcounter{equation}{0}

The computation of the effective potential in Section~\ref{sec:nobulkmass} can be explicitly carried through for the case when the back-reaction of the metric is small, that is when we expand around the usual AdS solution ($\delta/\chi \ll 1$). In this case the expanded bulk solutions are
\begin{eqnarray}
A(y)&=& ky + \frac{\delta^8}{16} \left( e^{8ky}-1 \right) + \order{\delta^9} \\
\phi(y) &=& \phi_0 + \frac{\sqrt{3}}{2 \kappa} \delta^4 e^{4ky} + \order{\delta^9}
\end{eqnarray} 
Solving the scalar BCs \eq{eq:BC2} for $\phi_0$ and $\delta$ at $\order{\delta^8}$, with the brane potentials \eq{eq:branepot}, we find,
\begin{eqnarray}
\phi_0 &=& \frac{v_1 (4 k - \lambda_0) \lambda_1 \chi^4 + v_0 (4 k + \lambda_1) \lambda_0 \mu_0^4}{(4 k - \lambda_0) \lambda_1 \chi^4 + (4 k + \lambda_1) \lambda_0 \mu_0^4} 
\label{phi0BC} \\
\delta^4 &=&  \frac{2 \kappa}{\sqrt{3}} (v_1 - v_0) \frac{\lambda_0 \lambda_1 \mu_0^4 \chi^4}{\mu_0^4 (4k + \lambda_1) \lambda_0 + \chi^4 (4 k - \lambda_0) \lambda_1}
\label{delta4BC}
\end{eqnarray}

For the effective potential we find at leading order in $\chi/\mu_0$:
\beq
V_{UV} = \Delta_0 \mu_0^4 + \order{\chi^8/\mu_0^4}
\eeq
\beq
V_{IR} = \left( \Delta_1 + \frac{4 k (v_0-v_1)^2 \lambda_1}{4k + \lambda_1} \right) \chi^4 + \order{\chi^8/\mu_0^4}
\eeq
where $\Delta_0 = \Lambda_0 - 6k/\kappa^2$ and $\Delta_1 = \Lambda_1 + 6k/\kappa^2$, the mistunings of UV and IR brane tensions. These expressions have the expected generic structure of (\ref{eq:VIR}-\ref{eq:VUV}). 

The rescaling of the potential after taking into account the dilaton-graviton mixing is given by (see Appendix~\ref{App:kinetic})
\beq
\frac{1}{K^2} = 1 + 2 \frac{\chi^2}{\mu_0^2}  \left( 1 +\frac{4 \kappa^2 \lambda_1^2 (v_0-v_1)^2}{3 (4k+\lambda_1)^2} \right)
+ \order{\chi^4/\mu_0^4}
\eeq
verifying that it is $\order{\chi^2/\mu_0^2}$.
 
\section{(Mistuned) Goldberger-Wise revisited\label{App:GW}}
\setcounter{equation}{0}

We revisit here the scenario discussed in Section~\ref{sec:CPR} under the assumption of a small mistune $\Delta_1$ and small back-reaction. This effectively corresponds to the analysis of GW supplemented by a small mistune ($\Delta_1\ll1$).

The solutions  to the bulk equations of motion close to AdS space, at $\order{\delta^8 \sim \kappa^2 \phi^2}$, and now also at $\order{\epsilon}$ and at all orders in $e^{\epsilon ky}$ is given by 
\begin{eqnarray}
A(y)&\simeq& ky + \frac{\delta^8}{16} \left( e^{(8-2\epsilon) ky}-1 \right) 
+ \frac{\kappa^2 \phi_0^2}{12} \left( e^{2\epsilon ky}-1 \right) 
+ \epsilon \frac{\delta^4 \kappa \phi_0 \sqrt{3}}{12} \left( e^{4 ky}-1 \right) 
\\
\phi(y) &\simeq& \phi_0 e^{\epsilon k y} + \frac{\sqrt{3}}{2 \kappa} \delta^4 e^{(4-\epsilon) ky}
\end{eqnarray}
where we have again fixed $A(0)=0$.
This expansion ensures that we include in the effective potential all terms at a given order up to $\order{\delta^8 \sim \kappa^2 \phi^2}$, including also $\order{\epsilon}$ terms, and at all orders in $e^{\epsilon ky}$.\\

Solving now the scalar BC's \eq{eq:BC2} for $\phi_0$ and $\delta$, with the brane potentials \eq{eq:branepot}, we find at $\order{\epsilon}$,
\begin{eqnarray}
\phi_0 &=& \frac{v_1 (4 k - \lambda_0) \lambda_1 \chi^{4-\epsilon} + v_0 (4 k + \lambda_1) \lambda_0 \mu_0^{4-\epsilon}}{(4 k - \lambda_0) \lambda_1 \chi^{4-2\epsilon} + (4 k + \lambda_1) \lambda_0 \mu_0^{4-2\epsilon}} 
+ \epsilon (...)
\label{phi0BCeps} \\
\delta^4 &\simeq&  \frac{2 \kappa}{\sqrt{3}} (\mu_0 \chi)^{4-\epsilon} \frac{\lambda_0 \lambda_1 (v_1 \mu_0^{-\epsilon} - v_0 \chi^{-\epsilon})}{\mu_0^{4-2\epsilon} (4k + \lambda_1) \lambda_0 + \chi^{4-2\epsilon} (4 k - \lambda_0) \lambda_1}
+ \epsilon (...)
\label{delta4BCeps}
\end{eqnarray}
where we have omitted   the $\order{\epsilon}$ terms to avoid clutter.
One can explicitly check that in the limit $\epsilon \to 0$ we recover \eq{phi0BC} and \eq{delta4BC}, furthermore in the limit $\lambda_0, \lambda_1 \to \infty$ limit we recover the well-known results of  Rattazzi and Zaffaroni~\cite{RZ}.

We can now compute the effective potential as in \eq{VUV} and \eq{VIR}, as a function of $\chi$, $\mu_0$, $\phi_0$, and $\delta$. The potential, using the expressions for $\phi_0$ and $\delta$ in \eq{phi0BCeps} and \eq{delta4BCeps}, at leading order in $\chi/\mu_0$, and in the limit $\lambda_0, \lambda_1 \to \infty$, is given by
\beq
V_{UV} = (\Delta_0 - \epsilon k v_0^2)  \mu_0^4
- \epsilon 2 k v_0 (\mu_0/\chi)^{2\epsilon} ( v_1-v_0 (\mu_0/\chi)^{\epsilon})  \chi^4
+ \order{\chi^8/\mu_0^4}
\eeq
\beq
V_{IR} = \left( \Delta_1 + 4 k (v_1-v_0(\mu_0/\chi)^{\epsilon})^2 
+ \epsilon (...)
\right) \chi^4 \left(1-\frac{\kappa^2(v_1-v_0(\mu_0/\chi)^{\epsilon})}{3} \right)+ \order{\chi^8/\mu_0^4}
\eeq
This is in agreement with our expectations: the value of $v_0$ is replaced by the running coupling $v_0 (\mu_0/\chi)^\epsilon$, making the dilaton quartic effectively run, which will allow a non-trivial minimum of the potential as in~\cite{GW,RZ}.


\section{Asymptotic matching\label{App:Matching}}
\setcounter{equation}{0}

In order to perform the asymptotic matching between the running and the boundary solutions it  is somewhat more convenient to change coordinates and rewrite the equations of motion (\ref{eq:bulkEOM}) for $\Phi(y)\equiv \phi(y/\epsilon)$ 
\begin{align}
&\epsilon\Phi^{\prime\prime}(y)-4A^\prime(y/\epsilon)\, \Phi^\prime(y)+4 k^2 \Phi(y) = 0\\
& A^{\prime\,2}(y/\epsilon)=\frac{\epsilon^2\kappa^2}{12}\Phi^{\prime\,2}(y)+\frac{\epsilon\kappa^2 k^2}{3}\Phi^2(y)+k^2
\end{align}
 where we have specified the bulk potential $V=-6k^2/\kappa^2-2\epsilon k^2\phi^2$.
These equations, for $\epsilon\ll 1$, show a boundary layer close to the IR brane, and one can directly apply the boundary layer theory of~\cite{Bender}. In the outer (UV) region where $\Phi$ and $A$ are slowly  varying we can neglect all the $\order{\epsilon}$ terms so that the solution is well approximated by
\begin{equation}
\Phi_r(y)=\Phi_0 e^{k y}\,,\qquad A_r^\prime=k
\end{equation}
where $\Phi_0$ is determined by the UV BC 
\begin{equation}
\Phi_0=v_0 e^{-k y_0}\,.
\end{equation}

As $\Phi$ approaches the boundary layer it starts running fast so that  we can neglect the mass term in the potential but not necessarily the cosmological constant contribution that can still be large: the boundary solution is thus given by
\begin{align}
\Phi_b(y)  =& \ \Phi_m - \frac{\sqrt{3}}{2 \kappa} \log \left( -\tanh \left[ 2k/\epsilon(y-\tilde{y}_c) \right] \right)\\
A^\prime_b(y/\epsilon) =& -k \coth \left( 4k/\epsilon(y-\tilde{y}_c) \right) 
\end{align}
 where we have defined $\tilde{y}_c=\epsilon y_c$.
The thickness of the boundary layer is  determined by 
\begin{equation}
y_{b}-\tilde{y}_c\sim \epsilon/2k
\end{equation} 
where $A'_b$ and $\Phi_b$ start approaching a constant. 

The IR BC fixes only one of the two integration constants; the other is going to be fixed by the asymptotic matching with the UV solution.

The asymptotic matching takes place at the edges of the inner and outer regions: in this overlap region both $\phi_{r,b}$ are solutions. For this boundary region we can take  for example  $y-\tilde{y}_c\sim \epsilon^{1/2}/2k$ for $\epsilon\ll 1$
\begin{equation}
\Phi_0 e^{k \tilde{y}_c}= \Phi_m\,.
\end{equation}
Of course we can  match in any other location as long it is in the overlapping region. It actually makes more sense to match at the branes at $y_1$ and $y_0$ which corresponds to taking $\epsilon$ small but finite
\begin{equation}
\Phi_0 e^{k y_1}= \Phi_m- \frac{\sqrt{3}}{2 \kappa} \log \left( -\tanh \left[ 2k/\epsilon(y_0-\tilde{y}_c) \right]\right)
\end{equation}
In this case the  full solution is given by
\begin{equation}
\Phi=\Phi_b+\Phi_r- \Phi_\mathrm{match}=\Phi_0 e^{k y}- \frac{\sqrt{3}}{2 \kappa} \log \left( \frac{\tanh \left[ 2k/\epsilon(\tilde{y}_c-y) \right]}{\tanh \left[ 2k/\epsilon(\tilde{y}_c-y_0) \right]} \right)
\end{equation}
where  $\Phi_0$ and $\tilde{y}_c$ in this expression are actually given in terms of $v_{0,1}$ and $y_{0,1}$ via the BCs.
 
 Notice that the scale $y_{br}$ where the back-reaction becomes important for the running solution is
 \begin{equation}
 y_{br}\sim \frac{1}{k}\log\left(\sqrt{3}/(\epsilon^{1/2}\kappa \Phi_0)\right)
 \end{equation}
 which agrees with QCD where $\Lambda_{QCD}$ is fixed by the UV coupling and its $\beta$-function:
 barring tuning, one expect $y_1\sim y_{br}\sim y_c$. 
 
 Transforming back to the original coordinates we obtain (\ref{eq:Phifull}).

\section{Dilaton kinetic term\label{App:kinetic}}
\setcounter{equation}{0}

We can parametrize the fluctuations of the 5D metric \eq{eq:ansatz} as
\beq
ds^2 = e^{-2 W(x,y)} \hat g_{\mu \nu}(x) d x^\mu d x^\nu - T(x,y)^2 dy^2 \ .
\label{eq:fluc}
\eeq
Upon evaluation of the 5D Ricci scalar term in \eq{eq:action} in terms of the metric \eq{eq:fluc}, one obtains
\begin{eqnarray}
\Lag_{eff}^{(kin)} &=& 
-\frac{1}{\kappa^2} \int_{y_0}^{y_1} dy \sqrt{g} \mathcal{R} \nn \\
&=& \frac{1}{\kappa^2} \int_{y_0}^{y_1} dy \sqrt{\hat g} e^{- 2 W} \left(
T \mathcal{R}[\hat g]
- 6 T (\partial W)^2
+ 6 \partial T \partial W
\right) \ ,
\label{eq:kinetic}
\end{eqnarray}
where we have dropped $\partial^\mu \hat g_{\mu \nu}$ terms by gauge-fixing the gravity fluctuations to be transverse.
The second line of \eq{eq:kinetic} makes evident that the 5D Ricci scalar contains both dilaton interactions (kinetic mixing) with gravity via the  first term in the parentheses, as well as the kinetic term for the dilaton, the second and third terms.

A proper parametrization to describe the the dilaton fluctuations is given by \cite{Csaki:2000zn}, 
\beq
W(x,y) = A(y) + F(x,y) \, , \quad T(x,y) = 1 + 2 F(x,y) \ .
\eeq
Expanding \eq{eq:kinetic} on the fluctuation, we obtain at quadratic order
\beq
\Lag_{eff}^{(kin)} = \frac{1}{\kappa^2} \int_{y_0}^{y_1} dy \sqrt{\hat g} e^{- 2 A(y)}
\left[
(1+2F^2) \mathcal{R}[\hat g] + 6 (\partial F)^2 + \order{F^3}
\right] \ .
\label{eq:kineticF}
\eeq

An ansatz for the case where we can neglect the fluctuations of the the bulk scalar, in particular for the limit $\lambda_0, \lambda_1 \to \infty$ in \eq{eq:branepot}, is given by $F(x,y) = f(x) e^{2A(y)}/e^{2A(y_1)}$. In this case one obtains
\beq
\Lag_{eff}^{(kin)} = \frac{1}{\kappa^2} \int_{y_0}^{y_1} dy \sqrt{\hat g} \left[
\left(e^{- 2 A(y)}+2f^2 \frac{e^{2 A(y)}}{e^{2A(y_1)}}\right) \mathcal{R}[\hat g] + 6 \frac{e^{2 A(y)}}{e^{2A(y_1)}} (\partial f)^2 + \order{f^3}
\right]
\eeq

\subsection{Dilaton reparametrizations}

In the main text we have parametrized the background dilaton solution by $\chi=e^{-A(y_1)}$ or in some cases it was more convenient to use a different parametrization  $\tilde{\chi} = e^{-ky_1}$.
Notice from \eq{eq:effpotentialy} that the latter does not yield automatically a quartic dilaton potential when the metric is not AdS.
In fact the relation between the two parametrizations can be explicitly computed for the metric \eq{eq:Aeps=0},
\beq
\chi^4 = \frac{2 \tilde \chi^4}{1+\tilde \chi^8 + (1-\tilde \chi^8) \cosh \left( \frac{2 \kappa}{\sqrt{3}} (v_1 - v_0) \right)} + \order{\tilde \chi^8/\tilde \mu_0^4}
\eeq
where $\tilde \mu_0 = e^{-ky_0}$ and where we have taken the limit $\lambda_0, \lambda_1 \to \infty$ for simplicity.
This nuisance is trivially solved by considering the dilaton interactions with gravity, of the form
\beq
\frac{1}{\kappa^2} \left( \int_{y_0}^{y_1} dy e^{- 2 A(y)} \right) \sqrt{\hat g} \mathcal{R}[\hat g]
= M_{Pl,eff}^2(y_0,y_1) \sqrt{\hat g} \mathcal{R}[\hat g]
\eeq
which defines the effective Planck scale, as function of $y_0$ and $y_1$. For a proper holographic interpretation of the effective dilaton potential it is convenient to factor out the dependence of $M_{Pl,eff}$ on $y_1$ and reabsorb it on the metric by a transformation $\sqrt{\hat g} \to \sqrt{\hat g}/K^2$, which will leave the gravitational kinetic term as $(\mu_0^2/\kappa^2 k) \sqrt{\hat g} \mathcal{R}[\hat g]$, with no dilaton contribution (interpreted as a purely elementary operator), and it will bring the dilaton potential to its expected $\chi^4 F(\chi/\mu_0)$ form, regardless of the identification of $\chi$ and $\mu_0$ as functions of $y_1$ and $y_0$ respectively. We can check this explicitly in the case at hand, where
\begin{eqnarray}
\frac{1}{k} \int_{y_0}^{y_1} dy e^{- 2 A(y)} &=& 
-\frac{1}{2} (\mu_0^2-\chi^2) + i \frac{\sqrt{2}}{8} \delta^2 \left[
\mathcal{B}\left(\left(\mu_0^4/\delta^4+\sqrt{1+\mu_0^8/\delta^8} \right)^2,3/4,1/2 \right)- \right. \nonumber \\
&& - \left. \mathcal{B}\left(\left(\chi^4/\delta^4+\sqrt{1+\chi^8/\delta^8} \right)^2,3/4,1/2 \right)
\right]
\end{eqnarray}
with $\mu_0$ and $\delta$ defined in \eq{eq:mu0def} and \eq{eq:deltadef} respectively, and $\mathcal{B}$ is the incomplete beta function.  Evaluating $\delta$ as obtained solving the $\phi$ BC's, \eq{eq:deltalambda}, one obtains the expected $K= 1 + \order{\chi^2/\mu_0^2}$. Likewise, if the dilaton is parametrized by $\tilde \chi$, the effective potential after the rescaling of the metric gets their expected scale invariant form,
\beq
\frac{e^{-4 A(y_1)}}{K^2} = \tilde{\chi}^4 \left( \mathrm{sech}^2\left( \frac{\kappa}{\sqrt{3}} (v_1 - v_0) \right) +\order{\tilde \chi^2/ \tilde \mu_0^2} \right) \ ,
\eeq
where the constant factor can be redefined away.

\section{Linear IR potential\label{App:Linearpotential}}
\setcounter{equation}{0}

Let us illustrate again our finding with a different IR brane potential, such that its contribution to the effective dilaton potential does vanish, contrary to the previous case.
\beq
V_1(\phi) = \Lambda_1 + \alpha_1 \phi \ .
\label{eq:branepotlin}
\eeq
and the same UV brane potential \eq{eq:branepot}.

The $\phi$ BCs fix now (again in the limit $\lambda_0 \to \infty$),
\begin{eqnarray}
\phi_0 &=& v_0 \mu_0^{\epsilon} \ , \\
\delta &=& \chi \left( \frac{4\sqrt{3}k-\sqrt{(4\sqrt{3}k)^2+ (\alpha_1 \kappa)^2}}{\alpha_1 \kappa} \right)^{1/4} \ .
\label{eq:solphiBCslin}
\end{eqnarray}

The effective potential,
\begin{eqnarray}
V_{UV} &=& \mu_0^4 \left[ 
\frac{6k}{\kappa^2}-\Lambda_0
\right] \ , 
\label{eq:UVpotlin} \\
V_{IR} &=& \chi^4 \left[ 
a_0 + \alpha_1 v_0 (\chi/\mu_0)^{-\epsilon}
 \right] \ , 
\label{eq:IRpotlin}
\end{eqnarray}
where
\beq
a_0 = \Lambda_1 + \frac{\sqrt{3}}{2 \kappa^2} \sqrt{(4\sqrt{3}k)^2+ (\alpha_1 \kappa)^2} 
+ \frac{\sqrt{3}}{2 \kappa} \alpha_1 \log \left[ \frac{4\sqrt{3}k}{\alpha_1 \kappa+\sqrt{(4\sqrt{3}k)^2+ (\alpha_1 \kappa)^2}} \right] \ .
\label{eq:a0}
\eeq
It is required that $a_0 > 0$ while $\alpha_1 <0$.

The minimum is found at
\beq
\frac{\vev{\chi}}{\mu_0} = \left( 
- \frac{a_0}{v_0 \alpha_1}
\right)^{-1/\epsilon} + \order{\epsilon} \ ,
\label{eq:minimumlin}
\eeq
while the dilaton mass
\beq
m_\chi^2 =  4 \epsilon a_0 \vev{\chi}^2 \ .
\label{eq:dilmasslin}
\eeq


\end{document}